\documentclass[aps,prb,reprint,groupedaddress,superscriptaddress,longbibliography]{revtex4-2}
\usepackage{amsmath}
\usepackage{amssymb}
\usepackage{graphicx}
\usepackage{subfigure}
\usepackage{comment}
\usepackage{ntheorem}
\usepackage{tabularx}
\usepackage{color}
\usepackage{bbm}
\usepackage[hidelinks]{hyperref}
\newcommand{\wsx}{\color {black}}
\newcommand{\zb}{\color {black}}

\begin{document}
\newtheorem{Definition}{Definition}[subsection]
   \title{  Super-enhanced Sensitivity in Non-Hermitian Systems at Infernal Points}
   \author{Shu-Xuan Wang}
   \email{wangshx65@mail.sysu.edu.cn}
   \affiliation{Guangdong Provincial Key Laboratory of Magnetoelectric Physics and Devices, State Key Laboratory of Optoelectronic Materials and Technologies, School of Physics, Sun Yat-sen University, Guangzhou 510275, China}
   \author{Zhongbo Yan}
   \email{yanzhb5@mail.sysu.edu.cn}
   \affiliation{Guangdong Provincial Key Laboratory of Magnetoelectric Physics and Devices, State Key Laboratory of Optoelectronic Materials and Technologies, School of Physics, Sun Yat-sen University, Guangzhou 510275, China}

   \date{\today}

   \begin{abstract}
      The emergence of exceptional points in non-Hermitian systems represents an intriguing phenomenon characterized by the coalescence of eigenenergies and eigenstates. When a system approaches an exceptional point, it exhibits a heightened sensitivity to perturbations compared to the conventional band degeneracy observed in Hermitian systems. This sensitivity, manifested in the splitting 
      of the eigenenergies, is amplified as the order of the exceptional point increases. Infernal points constitute a unique subclass of exceptional points, distinguished by their order escalating with the expansion of the system's size. 
      In this paper, we show that, when a non-Hermitian system is at an infernal point, 
      a perturbation of strength $\epsilon$, which couples the two opposing  
      boundaries of the system, causes the eigenenergies to split according to  the law
      $\sqrt[k]{\epsilon}$, where $k$ is an integer proportional to the system's size.
      Utilizing the perturbation theory of Jordan matrices, we demonstrate that the exceptional sensitivity 
      of the eigenenergies at infernal points to boundary-coupling perturbations is a ubiquitous phenomenon, irrespective of the specific form of the non-Hermitian Hamiltonians. Notably, we find that this phenomenon 
      remains robust even when the system deviates {\zb substantially} from the infernal point. 
      The universal nature and robustness of this phenomenon suggest potential applications in enhancing sensor sensitivity. 
   \end{abstract}

   \maketitle

   \emph{Introduction}---The hermiticity of Hamiltonians is a fundamental assumption 
   in quantum mechanics, yet it has been shown that non-Hermitian Hamiltonians can offer 
   an effective description to many open systems~\cite{PhysRevLett.70.2273, rotter2009non}. 
   In recent years, a wide array of exotic phenomena stemming from non-hermiticity have been 
   successively unveiled, sparking increased interest in the study of non-Hermitian physics. 
   Among the myriad of phenomena observed, the emergence of exceptional points (EPs)~\cite{W.D.Heiss_2004, berry2004physics}, which 
   signify the coalescence of eigenenergies and eigenstates of the non-Hermitian Hamiltonians, has garnered immense attention 
   due to its potential applications across various fields~\cite{PhysRevLett.120.146402, PhysRevLett.123.066405, PhysRevResearch.4.L022064, PhysRevLett.124.186402, PhysRevLett.126.086401, PhysRevLett.127.196801, PhysRevB.108.085104, PhysRevResearch.5.L042010, PhysRevLett.118.045701, PhysRevB.99.081102, PhysRevB.99.161115, PhysRevB.97.075128, PhysRevB.99.041202, PhysRevLett.128.226401, PhysRevLett.127.186602, gohsrich2024, PhysRevResearch.6.043023, PhysRevX.6.021007, Zhou:19,PhysRevLett.86.787, PhysRevLett.101.080402,PhysRevLett.103.134101, Zhu:10, pnas.1603318113, regensburger2012parity, zhen2015spawning, science.aap9859, cerjan2019experimental,PhysRevX.6.021007, science.abd8872,PhysRevLett.123.237202,PhysRevLett.104.153601}. A particularly intriguing property 
   of EPs is their response to perturbations. Specifically, for a $k$-th order EP, where $k$ 
   represents the number of eigenenergies and eigenstates that coalesce into one, it has been demonstrated that, 
   under the influence of a perturbation with a strength proportional to $\epsilon$, 
   the splitting of eigenenergies can be proportional to $\sqrt[k]{\epsilon}$. In contrast to the splitting 
   of band degeneracy in Hermitian systems, which scales directly with
   $\epsilon$, it is evident that  high-order EPs,  in particular,
   exhibit a much greater sensitivity to infinitesimally small perturbations. 
   This fascinating property is believed to offer significant potential for 
   enhancing sensor sensitivity, prompting a range of theoretical and experimental 
   investigations~\cite{PhysRevLett.112.203901, PhysRevA.93.033809, chen2017exceptional, hodaei2017enhanced, PhysRevA.98.023805, PhysRevLett.123.213901, PhysRevLett.122.153902, PhysRevApplied.12.024002, PhysRevLett.125.240506, PhysRevA.101.053846, PhysRevResearch.6.033284}.

   Recently, a unique subclass of EPs, referred to as infernal points (IPs), 
   has been uncovered in non-Hermitian lattice systems subject to open boundary conditions (OBC)
   through the framework of non-Bloch band theory~\cite{denner2021exceptional, 
   denner2023infernal, PhysRevB.105.075420, wang2024infernalpoints}. At an IP, all eigenstates of the entire system 
   coalesce into a limited number of eigenstates~\cite{wang2024infernalpoints}. In essence, an IP represents a high-order EP, with its order 
   directly proportional to the number of lattice sites. This characteristic is particularly intriguing because 
   it suggests that the order of this class of EPs can be readily incremented by adjusting the system's size, 
   in stark contrast to the conventional method of achieving higher-order EPs, which demands precise parameter tuning under stringent symmetry requirements{\zb ~\cite{Mandal2021,Zhang2020HOEP}}. As the order is macroscopically large, it is natural to anticipate that an IP will 
   response to a minor perturbation in an exceptionally sensitive manner. However, it is crucial to recognize that not all types of perturbations will result in the same order of splitting, especially for higher-order EPs. Therefore, gaining a general understanding of the conditions required for a perturbation to induce the most pronounced splitting of IPs is of fundamental importance 
   for their in-depth exploration and potential future applications.
   
   {\zb In this work, we first employ the one-dimensional (1D) Hatano-Nelson (HN) model 
   to show that boundary-coupling perturbations induce maximal spectral splitting 
   when the system is at an IP. Since non-Hermitian Hamiltonians
   at IPs inherently adopt a Jordan block structure, we derive general conditions under which perturbations cause the most significant splitting, leveraging Jordan matrix perturbation theory. This framework directly explains the extreme 
   sensitivity of non-Hermitian systems at IPs to boundary-coupling perturbations. 
   Remarkably, we find that this high sensitivity persists even when the system deviates substantially from the IP. 
   This discovery significantly reduces the challenging for the experimental implementation and practical 
   applications of the predicted phenomenon. }

\begin{figure}
       \centering
       \subfigure{\includegraphics[scale=0.33]{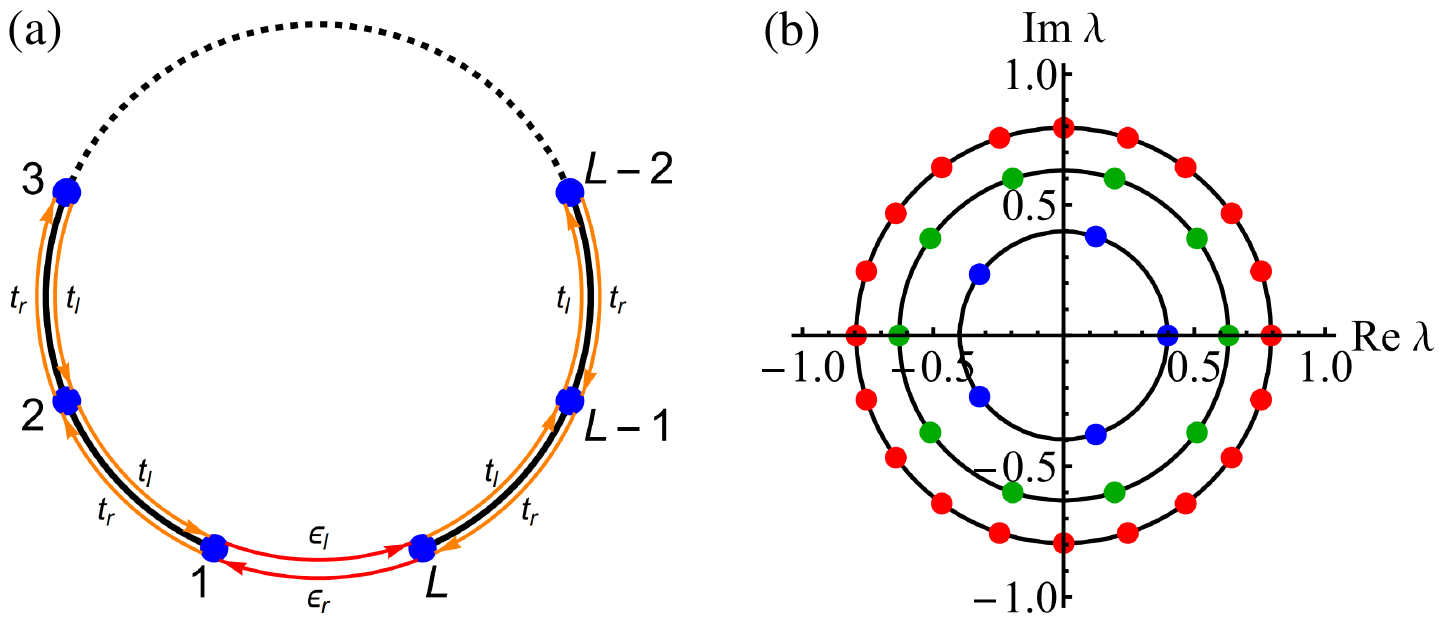}}
       \caption{(a) A schematic diagram of the HN model under a boundary-coupling perturbation. $\epsilon_{r}$
       and $\epsilon_{l}$ characterize the strength of the perturbation.  
       (b) The blue, green and red dots depict eigenenergies for $L=5$, $10$ and $20$, respectively. 
       From the innermost to the outermost, the radius of each circle corresponds to $\sqrt[5]{10^{-2}}$, $\sqrt[10]{10^{-2}}$ and $\sqrt[20]{10^{-2}}$, respectively. The parameters used in (b) are $t_{l}=1$, $t_{r}=0$, and $\epsilon_{r}=\epsilon_{l}=10^{-2}$.}
       \label{fig1}
     \end{figure}

   \emph{Insights from the HN model}---We begin with the 1D non-Hermitian 
   HN model~\cite{PhysRevLett.77.570}, which, under OBC, is expressed as follows:
     \begin{equation}
        H_{HN} = \sum_{n=1}^{L-1} t_r c_{n+1}^{\dagger} c_n  + t_l c_{n}^{\dagger} c_{n+1},
        \label{1}
     \end{equation}
    where $t_r$ and $t_l$ represent the hopping amplitudes, and $L$ denotes the total number of 
    lattice sites, as depicted in Fig.~\ref{fig1}(a). In our previous work~\cite{wang2024infernalpoints}, we demonstrated that an IP emerges when either $t_r = 0$ or 
    $t_{l}=0$, indicating unidirectional hopping.  Specifically, when $t_r = 0$, 
    {\zb the HN Hamiltonian adopts a Jordan block structure, and}
    all eigenstates of the system converge to a single eigenstate of the form $\Phi_{0} = (1,0,0,\cdots,0)^T$, accompanied by the covergence of all eigenenergies to a single energy level at $E=0$. 

    When the system is at this IP, how do the eigenenergies respond to a perturbation?  
    We find that the impact of the perturbation on the eigenenergies depends on its specific form. 
    For an on-site perturbation of the form $H^{\prime} = \epsilon c_{n}^{\dagger} c_n$ with $\epsilon \ll t_l$, we find the following: (1) When 
    $n=1$, the system retains its single eigenstate $\Phi_{0} = (1,0,0,\cdots,0)^T$, and the sole 
    eigenenergy is shifted to $E=\epsilon$. (2) When $n\in [2,L]$, there are two eigenenergies: one at 
    $E=0$ and the other at $E=\epsilon$. The corresponding eigenstates are $\Phi_{0} = (1,0,0,\cdots,0)^T$
    and $\Phi_{\epsilon} = (1,\epsilon/t_{l},...,(\epsilon/t_{l})^{n-1},0,\cdots,0)^T$, respectively. 
    Apparently, the on-site potential perturbation cannot result in the order of splitting desired. 
  
    The HN model is renowned for exhibiting non-Hermitian skin effect~\cite{PhysRevLett.121.086803}, which notably 
    results in a significant disparity between the energy spectrum observed under periodic boundary conditions (PBC) 
    and that observed under OBC. Having observed this fact, we introduce a perturbation that connects the left and right ends of the chain, as depicted in Fig.~\ref{fig1}(a). Specifically, the perturbation takes the form:
     \begin{equation}
       H_p = \epsilon_{r}c_{1}^{\dagger} c_L +   \epsilon_{l}c_{L}^{\dagger} c_1,
       \label{2}
     \end{equation}
   where $\epsilon_{r}$ and $\epsilon_{l}$ represent the hopping strength between the two ends of the chain. Accordingly, 
   the matrix form of the Hamiltonian describing the perturbed system at  $t_r = 0$ reads 
     \begin{equation}
        H = 
         \begin{pmatrix}
            0 & t_l &    &    & \epsilon_{r}  \\
            0 & 0  & t_l &  &            \\
              & \ddots  &  \ddots & \ddots &  \\
              &         &   0   &  0  & t_l   \\
            \epsilon_{l}   &     &     &  0 & 0
         \end{pmatrix}_{L \times L}.
         \label{3}
     \end{equation}
   The characteristic polynomial of $H$ is {\zb (see details in Sec. I of the Supplemental Material (SM)~\cite{supplemental})},
     \begin{equation}
        \begin{split}
           f(\lambda) & = \det \left[H - \lambda \mathbb{I}_{L \times L}  \right] \\
                      & = (-\lambda)^{L} + (-1)^{L-1} \epsilon_{l} t_{l}^{L-1} - \epsilon_{r}\epsilon_{l} (-\lambda)^{L-2}.
        \end{split}
        \label{4}
     \end{equation}
     It is readily apparent that $\epsilon_{r}$ and $\epsilon_{l}$ have remarkably different effects 
     on the eigenenergies. First, consider the scenario where $\epsilon_{l}=0$ but $\epsilon_{r}\neq0$. 
     Solving $f(\lambda)=0$ in this case immediately yields $\lambda=0$, indicating that all eigenenergies remain degenerate. 
     Conversely, if $\epsilon_{r}=0$ but $\epsilon_{l}\neq0$, then one can find 
     $L$ roots that satisfy $f(\lambda)=0$. These roots can be compactly expressed as
     \begin{equation}
       \lambda_p =\left( \frac{\epsilon_{l}}{t_l} \right)^{\frac{1}{L}} t_l e^{i \frac{2 \pi p}{ L}},
       \qquad p = 1, 2, \cdots, L.
       \label{5}
     \end{equation}
  Intriguingly, these $L$ roots are evenly distributed along the circumference of a circle{\zb ~\cite{Guo2024}}, with the radius of this circle representing the extent of the splitting of the IP.  
  Given that the radius is equal to $t_{l}\sqrt[L]{\epsilon_{l}/t_{l}}$, 
  it becomes evident that the eigenenergy splitting is directly proportional to  $\sqrt[L]{\epsilon_{l}}$. 
  This signifies that the splitting, which is induced by the perturbation $\epsilon_{l}$, 
  amplifies with the growth of the system's size. However, it is important to note that 
  the splitting does not diverge, rather, it saturates at $t_{l}$ in
  the limit of large $L$. This result admits a physical interpretation: in the infinite system size limit, an infinitesimal coupling that slightly modifies the OBC to approach PBC will restore the eigenenergies predicted by Bloch band theory. Indeed, consider the scenario where $\epsilon_{l}=t_{l}$, which corresponds to the system adopting PBC. 
  In this case, Bloch band theory predicts that all eigenenergies will have the exact form given in Eq.~(\ref{5}). 
  
  We have demonstrated above  that $\epsilon_{r}$ does not influence the IP when  $\epsilon_{l}$ is equal to zero. 
  When both $\epsilon_{l}$ and $\epsilon_{r}$ are nonzero but much smaller than $t_{l}$, 
  the splitting should be primarily  caused by $\epsilon_{l}$. To illustrate this, Fig.~\ref{fig1}(b) displays the eigenenergy distributions obtained by numerically diagonalizing the Hamiltonian given in Eq.~(\ref{3}) under the condition $\epsilon_{r}=\epsilon_{l}=\epsilon$. The numerical results align well with the analytical solutions presented in Eq.~(\ref{5}), confirming that $\epsilon_{l}$ plays the dominant role in inducing the splitting.

   \par

   \emph{The generality of the phenomenon}---We have demonstrated through the 
   HN model that the eigenenergies at an IP exhibit exceptional sensitivity to boundary-coupling perturbations. 
   This naturally leads to the question: Is this phenomenon a universal occurrence, or is it specific to particular models? 
   {\zb Observing that non-Hermitian Hamiltonians at IPs inherently adopt a Jordan block structure, we employ the perturbation theory of Jordan matrices~\cite{seyranian2003multiparameter, lidskii1966perturbation, MA199845, MIVishik_1960, doi:10.1137/S0895479895294666} to 
    demonstrate the generality of the phenomenon. At the same time, we establish a universal  condition 
    under which a perturbation of strength $\epsilon$ guarantees the spectral splitting scales as $\sqrt[L]{\epsilon}$.}

    Consider a general 1D non-Hermitian lattice system of length $L$, where the Hamiltonian at an IP is denoted by $H_{\rm IP}$. 
    Without loss of generality, assume that $H_{\rm IP}$ has a right eigenstate  $|u_0\rangle$ (which is not a topological mode) at an energy 
    $E=E_{\rm IP}$, with its corresponding left eigenstate being $|v_0\rangle$. Since the
   emergence of an IP stems from the collapse of the generalized Brillouin zone~\cite{wang2024infernalpoints}, all 
    eigenstates of a band will coalesce to one at the IP. Therefore, $E_{\rm IP}$
    can also be viewed as a nonderogatory eigenenergy{\wsx \footnote{nonderogatory eigenenergy here means that the geometric multiplicity of this eigenenergy is one.}} of $H_{\rm IP}$~\cite{seyranian2003multiparameter}  whose {\wsx algebraic} multiplicity is 
    $k=L-\alpha$. Here $k$ represents the number of eigenstates within a band, while 
    $\alpha$ is a finite integer that counts the number of topological modes in the system, 
    which is independent of the system's size $L$.   

    Next consider the presence of a perturbation $H_{\rm pert} = \epsilon H_1$ with  $\epsilon \ll 1$ being a dimensionless parameter. 
    The perturbation theory of Jordan matrix \cite{seyranian2003multiparameter, lidskii1966perturbation, MA199845, MIVishik_1960, doi:10.1137/S0895479895294666} tells that, if $\langle v_0 | H_1 | u_0 \rangle=\Lambda$ is nonzero, then the perturbed 
    eigenenergies and eigenstates with the leading-order corrections are given by (a physical approach to the case 
    when $H_1$ only contains hoppings between the two ends of the system is provided in {\zb Sec. II of the} SM~\cite{supplemental})
     \begin{gather}
      E_{p,i} = E_{\rm IP} + \epsilon^{\frac{1}{k}} e_i + o(\epsilon^{\frac{1}{k}}) \qquad (i=1,2,\cdots,k),
      \label{7}
      \\
      |u_{p,i}\rangle = |u_0\rangle +  \epsilon^{\frac{1}{k}} e_i |u_1\rangle + o(\epsilon^{\frac{1}{k}}) \quad (i=1,2,\cdots,k),
      \label{8}
     \end{gather}
   where $e_i$ with $i=1,2,\cdots,k$ denotes the $i$-th root of the equation $e - \Lambda^{\frac{1}{k}} = 0$;
   $|u_{p,i}\rangle$ is the eigenstate associated with the eigenenergy $E_{p,i}$, and $|u_1\rangle$
   satisfies  $(H_{\rm IP} - E_{\rm IP}) |u_1\rangle = |u_0\rangle$. 
   Eqs.~\eqref{7} and \eqref{8} reveal that if $\Lambda$ is nonzero, the perturbation $H_{\rm pert}$ will split the coalescence and recover the existence of  $k$ eigenstates,  and importantly, the splitting of eigenenergies is proportional to $\sqrt[k]{\epsilon}\sim \sqrt[L]{\epsilon}$ when 
   $L$ is much larger than $\alpha$. {\wsx Since $\lim_{L \rightarrow \infty } \sqrt[L]{\epsilon} = 1$ for $\forall \epsilon >0$, a target signal inducing an arbitrarily weak boundary-coupling perturbation to non-Hermitian systems at an IP 
   will become detectable in the limit of sufficiently large system size $L$. 
   This sensitivity scaling far exceeds that achievable with conventional EPs~\cite{PhysRevLett.112.203901,hodaei2017enhanced},
   suggesting that IPs could enable super-enhanced sensor designs through their unique spectral properties. }

   In Ref.~\cite{wang2024infernalpoints}, it has been shown that all right eigenstates at an IP are localized at finite sites near the boundary of the system, and similarly, all corresponding left eigenstates are extremely localized at the opposite boundary. Hence, $|u_0\rangle$ and $|v_0\rangle$ are localized at opposite ends ({\zb see more discussions in Sec. II of the SM~\cite{supplemental}}). To make $\langle v_0 | H_1 | u_0 \rangle$ nonzero,  it is evident that a natural and physical choice of the perturbation  $H_1$ is
    to contain the hopping between the two ends of the system,
   as previously illustrated by the HN model. Since all results in this section are independent of the specific form of $H_{\rm IP}$, it 
   suggests that the eigenenergies of any 1D non-Hermitian lattice system at an IP will always be extremely sensitive to hoppings between 
   the two ends of the system. {\zb In Sec. III of the SM~\cite{supplemental}, 
   we demonstrate the model-independence of this phenomenon  using a completely different model.} 
   
   {\wsx Before concluding this section, we compare our results with the non-Hermitian topological sensor proposed in Ref.~\cite{PhysRevLett.125.180403}. Focusing on topological zero modes, the authors showed via perturbation theory 
   that a boundary-coupling perturbation induces an exponential energy shift of the form:
   $\Delta E\approx\epsilon \kappa e^{\alpha L}$ for $L\gg1$, where $\kappa$ and $\alpha$ are model-dependent parameters. 
   Although this sensitivity also grows exponentially with system size, the scaling $\Delta E\sim \epsilon$ (for fixed 
    $L$) contrasts sharply with the $\sqrt[L]{\epsilon}$ dependence, indicating fundamentally distinct underlying mechanisms.
   Moreover,  the formula $\Delta E\approx\epsilon \kappa e^{\alpha L}$ is only physically meaningful below a critical system size, 
   as it would otherwise diverge unphysically as $L\rightarrow+\infty$. In contrast, 
   these formulas in Eqs.~(\ref{7}) and (\ref{8}) remain valid for arbitrarily large  $L$, since
   $\lim_{L \rightarrow \infty } \sqrt[L]{\epsilon} = 1$. The most significant 
   distinction lies in the scope of description: while the formula developed in Ref.~\cite{PhysRevLett.125.180403} specially 
   addressed the behavior of topological zero modes, our framework provides a complete characterization encompassing all eigenstates except the topological modes through Eqs.~(\ref{7}) and (\ref{8}). This demonstrates that IP-based sensitivity is inherently a collective phenomenon, 
   offering the practical advantage of signal detection without requiring mode-specific 
    measurements---a substantial simplification for potential implementations. }

   \emph{Robustness of the phenomenon}---While the special property 
   of IPs can result in super-enhanced sensitivity to perturbations fulfilling the condition discussed above, 
   the occurrence of IPs typically hinges on a precise tuning of the system's parameters~\cite{wang2024infernalpoints}. 
   If this super-enhanced sensitivity were confined solely to systems exactly at an IP, its practical applications would be severely limited. 
   Therefore, it is crucial to investigate whether this heightened sensitivity persists even when the system deviates from the exact IP. 
   Fortunately, our findings affirm that it does, highlighting the robustness of this phenomenon.
  
\begin{figure}
      \centering
      \subfigure{\includegraphics[scale=0.3]{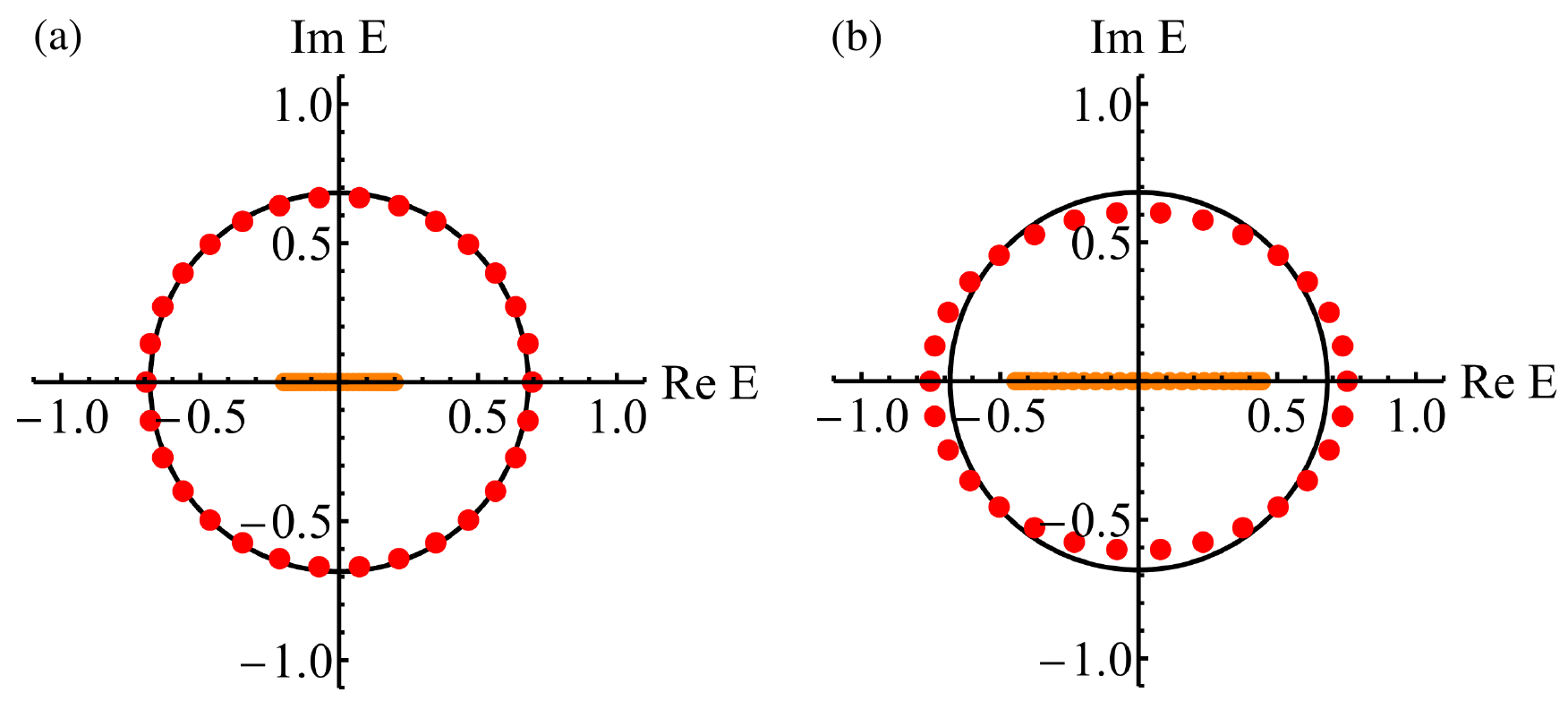}}
      \caption{The spectrum of the HN model away from the IP. The orange dots depict 
      eigenenergies without considering the boundary-coupling perturbation, while the dots in 
      red depict eigenenergies in the presence of the perturbation. We set $\epsilon_{r}=\epsilon_{l}=\epsilon$,
      and the radius of the black circle is given by $\sqrt[L]{\epsilon}$ with $L=30$ and $\epsilon = 10^{-5}$. 
      In panel (a), $t_{l}=1$, and $\delta = 0.01$.  In panel (b), $t_{l}=1$, and $\delta = 0.05$.}
    \label{fig2}
    \end{figure}

   Again we use the HN model to demonstrate the robustness of the phenomenon. Now we set $t_l = 1$ and $t_r = \delta \ll 1$, so 
   that the system deviates from the precise IP at $t_{r}=0$. 
   According to the non-Bloch band theory~\cite{PhysRevLett.123.066404}, the spectrum for a system with length $L$ subject to OBC is given by $E(\delta,q) = 2 \sqrt{\delta} \cos q $, where 
   $q=0,2\pi/L,...,2\pi(L-1)/L$. It is evident that the band width is exclusively determined by $\delta$ and is independent of $L$. 
   The numerical results presented in Fig.~\ref{fig2} show that the eigenenergies remain exceptionally sensitive 
   to the boundary-coupling perturbation given in Eq.~\eqref{2} even though the system deviates from the precise IP.
   Particularly, we find that the perturbed eigenenergies are still approximately equal to $\sqrt[L]{\epsilon}$ even if $\epsilon \ll \delta$ 
   ($\epsilon_{r}=\epsilon_{l}=\epsilon$ is set), 
   demonstrating the robustness of the phenomenon.

   To understand the robustness of the phenomenon uncovered above, we employ the perturbation theory again. 
   To be specific, we start with $H_{\rm IP}$ and treat the hopping {\wsx leading to the deviation} of the system from 
   the IP as a perturbation, $H_{\delta} = \delta H_2$, where $\delta \ll 1$ is also a dimensionless parameter and $H_2$ preserve 
   the translation symmetry within the open-boundary system (the hopping range of the terms in $H_{\delta}$ are assumed to be finite). 
   Before adding the perturbation $H_{\rm pert}$, the whole system is described by $H_{\rm IP} + H_{\delta}$. Based on the non-Bloch
   band theory~\cite{PhysRevLett.123.066404,wang2024infernalpoints}, the generalized Brillouin zone of this system 
   will be split from a point to a closed curve, and the spectrum satisfies
     \begin{equation}
       \left[ E (\beta (\delta)) - E_{\rm IP}  \right] \propto \delta^{r},
       \label{9}
     \end{equation}
   where $\beta (\delta)$ characterizes the generalized Brillouin zone, and $r > 0$ is a finite rational number, 
   which is independent of the system size. Take the HN model as an example. If we choose $H_{\rm IP} = \sum_{n}  c_{n-1}^{\dagger} c_{n}$ and $H_{\delta} = \delta \sum_{n} c_{n+1}^{\dagger} c_n$. In this case, $E_{\rm IP} = 0$, $E (\beta (\delta)) - E_{\rm IP} = 2 \sqrt{\delta} \cos q \propto \sqrt{\delta}$, indicating $r =1/2$. 

   To assess the impact of the perturbation $H_{\rm pert}$, it is important to first recognize that 
   $H_{\rm pert}$ should not be considered as a minor disturbance to  $H_{\rm IP} + H_{\delta}$. To illustrate this point, let's once again consider
   the HN model as an example. For the total Hamiltonian $H_{\rm IP} + H_{\delta}$, 
   the right eigenstate corresponding to $E (\beta (\delta))$ is $| u(\beta(\delta)) \rangle = \frac{1}{\sqrt{L}} (1, \beta, \beta^2, \cdots, \beta^{L-1})^T$ and its left eigenstate is $\langle v(\beta(\delta)) |= \frac{1}{\sqrt{L}} (1, \beta^{-1}, \beta^{-2}, \cdots, \beta^{-(L-1)})$. 
   The left and right eigenstates satisfy $\langle v(\beta(\delta)) | u(\beta(\delta)) \rangle = 1$. Employing the perturbation 
   theory, the correction to $E (\beta (\delta))$ induced by $H_p$ given in Eq.~\eqref{2} is 
   \begin{eqnarray}
    \Delta E(\beta (\delta)) &\approx& \langle v(\beta(\delta)) |H_p| u(\beta(\delta)) \rangle  \nonumber\\
    &=& \frac{1}{L} (\epsilon_{r}\beta^{L-1} + \epsilon_{l}\beta^{-(L-1)}).
   \end{eqnarray}
   {\zb Since $|\beta (\delta)|$ is smaller than 1 for $\delta<1$, 
   $\Delta E(\beta (\delta))$ always goes divergent as $L$ goes to infinity. }
    This result indicates that the eigenenergies of this system is more sensitive to the hopping between 
    the two ends than other types of perturbations. 
    The divergence of $\Delta E(\beta (\delta))$ suggests that treating $H_{p}$ as a perturbation to $H_{\rm IP} + H_{\delta}$ cannot 
    yield the accurate spectrum. Given that $H_{p}$ exerts a greater influence on the spectrum compared to $H_{\delta}$, 
    it is more appropriate to treat $H_{\delta}$ as a perturbation  to $H_{\rm IP} + H_{p}$. 
    
    Based on the eigenenergies and eigenstates of $H_{\rm IP} + H_{\rm pert}$ given in Eqs.~\eqref{7} and \eqref{8}, 
    the eigenenergies, after taking into account  the first-order corrections induced by $H_{\delta}$,  are modified as
    $\tilde{E}_{p,i} \approx E_{p,i} + \langle v_{p,i} | H_{\delta} | u_{p,i}   \rangle  $, where $| v_{p,i} \rangle$ is the 
    corresponding left eigenstate of $| u_{p,i} \rangle$ and $\langle v_{p,i} | u_{p,j} \rangle =\delta_{ij}$~\footnote{After 
    including the perturbation  $H_{\rm pert}$, the eigenenergy splits, and $E_{p,i}$ is no longer a nonderogatory eigenenergy but becomes a simple eigenenergy. Thus $|v_{p,i}\rangle \not= |v_0\rangle +  \epsilon^{\frac{1}{k}} e_i^{*} |v_1\rangle + o(\epsilon^{\frac{1}{k}})$}. Write down the result explicitly,  
    \begin{equation}
         \tilde{E}_{p,i} \approx  E_{\rm IP} + \epsilon^{\frac{1}{k}} e_i + \delta  \langle v_{p,i} | H_2 | u_{p,i} \rangle .
       \label{10}
     \end{equation}
    When the system's size is sufficiently large, {\zb the second term on the right-hand side of 
    Eq.~(\ref{10}) can dominate the third term over a broad range of $\delta$. 
    This result suggests that the exceptional sensitivity of eigenenergies to $H_{\rm pert}$ 
    can persist even when the system deviates substantially from the IP, as illustrated in Fig.~\ref{fig2}(b). }

   \emph{Discussions and conclusions}---The IP that arises in open-boundary non-Hermitian systems 
   offers a versatile route to achieving EPs of arbitrarily high order through mere adjustment of the 
   system's size. When such a system is at an IP, we 
  have shown that a perturbation of strength $\epsilon$, which couples the two opposing  
  boundaries of the system, causes the eigenenergies to split according to  the law
  $\sqrt[k]{\epsilon}$, where $k$ is proportional to the system's length and represents the order of the EP.
  Utilizing the perturbation theory of Jordan matrices, we have demonstrated that the sensitivity 
  of the eigenenergies to this type of boundary-coupling perturbation is a ubiquitous phenomenon. Moreover, based on 
  both perturbation theory and exact numerical calculations, we have confirmed that this phenomenon 
  remains robust even when the system deviates substantially from the IP.

  The observation of the phenomenon predicted in this paper is 
  within the current state-of-the-art experimental conditions. 
  Specifically, the sensitivity of the eigenenergies to boundary-coupling perturbations
  stems from the significant discrepancy in the spectra of non-Hermitian systems under 
  OBC and PBC. This sensitivity makes it possible to observe a strong response of the eigenenergies to perturbations in 
  systems that exhibit strong non-Hermitian skin effects. Notably, such systems have been implemented across a diverse range of flexible platforms, including metamaterials~\cite{Zhang2021NHSE}, cold-atom systems~\cite{PhysRevLett.129.070401,zhao2025two}, electric circuits~\cite{PhysRevResearch.2.023265,PhysRevB.105.195131,zou2021observation,PhysRevApplied.22.L031001}, phononic crystals~\cite{zhou2023observation}, acoustic systems~\cite{zhang2021acoustic,PhysRevLett.130.017201} and others~\cite{PhysRevLett.131.207201}. 
  On the basis of these systems implemented in experiments, we anticipate 
  that the super-enhanced sensitivity of the eigenenergies to perturbations 
  will be observed as the system is designed to approach the IP.

  In summary, the eigenenergies of non-Hermitian systems at IPs collectively exhibit exceptional sensitivity 
  to boundary-coupling perturbations. This unique property holds
  great potential for applications involving sensitive measurements.

   \emph{Acknowledgments}--We thank Zheng-Yang Zhuang and Yao Zhou for helpful discussions. This work is supported by the National Natural Science Foundation of China (Grant No. 12174455) and Guangdong
   Basic and Applied Basic Research Foundation (Grant No.
   2023B1515040023).

   \bibliography{sensitivity}

   \clearpage

   \onecolumngrid
    \begin{center}
      \Large{Supplemental Material for ``Super-enhanced Sensitivity in Non-Hermitian Systems at Infernal Points''}
    \end{center}
 
   \renewcommand\thesection{\Roman{section}}

   \section{Detailed derivation of Eq.~(4)} 
      For the Hamiltonian given in Eq.(3), its characteristic polynomial is determined by calculating the determinant of the 
      following matrix, 
        \begin{equation}
          H - \lambda \mathbb{I}_{L \times L} = 
            \begin{pmatrix}
               -\lambda & t_l &    &    & \epsilon_{r}  \\
               0 & -\lambda  & t_l &  &            \\
                 & \ddots  &  \ddots & \ddots &  \\
                 &         &   0   &  -\lambda  & t_l   \\
               \epsilon_{l}   &     &     &  0 & -\lambda
            \end{pmatrix}_{L \times L} .
            \tag{S1}
            \label{S1}
        \end{equation}
      By expanding the determinant according to the first column,
        \begin{equation}
           \begin{split}
              f (\lambda ) =& \det \left[H - \lambda \mathbb{I}_{L \times L}  \right]\\
              =& -\lambda \det \left[ 
                \begin{pmatrix}
                  -\lambda & t_l &    &    &  \\
                  0 & -\lambda  & t_l &  &            \\
                    & \ddots  &  \ddots & \ddots &  \\
                    &         &   0   &  -\lambda  & t_l   \\
                    &     &     &  0 & -\lambda
                \end{pmatrix}_{(L-1) \times (L-1)}   
                \right]
                \\
                &+ (-1)^{L-1} \epsilon_{l} \det \left[
                  \begin{pmatrix}
                     t_l & 0 &  & & \epsilon_{r}  \\
                     -\lambda & t_l & 0 & & \\
                     & \ddots & \ddots & \ddots & \\
                     &  &  \ddots & \ddots & 0  \\
                     &  &  &  -\lambda & t_l
                  \end{pmatrix}_{(L-1) \times (L-1)}
               \right]
               \\
               =& -\lambda f_1 + (-1)^{L-1} \epsilon_{l} f_2 .
           \end{split}
           \tag{S2}
            \label{S2}
        \end{equation}
      Apparently, $f_1 = (-\lambda)^{L-1}$ and 
        \begin{equation}
           \begin{split}
               f_2 =& t_l \det  \left[
                 \begin{pmatrix}
                    t_l &  &  &   \\
                    -\lambda & t_l & & \\
                    &  \ddots  & \ddots &  \\
                    &  &  -\lambda & t_l 
                 \end{pmatrix}_{(L-2) \times (L-2)}
                 \right]
                 \\
                 & + (-1)^{L-2} \epsilon_{r} \det \left[
                    \begin{pmatrix}
                       -\lambda & t_l &  &  \\
                       &  \ddots  & \ddots & \\
                       &    &  \ddots  & t_l \\
                       &  &  & -\lambda 
                    \end{pmatrix}_{(L-2) \times (L-2)}
                    \right]
                  \\
                  =& t_l^{L-1} + (-1)^{L-2} (-\lambda)^{L-2} \epsilon_{r}.
           \end{split}
           \tag{S3}
           \label{S3}
        \end{equation}
     Therefore, one finds 
        \begin{equation}
           \begin{split}
              f (\lambda) &= (-\lambda)^L + (-1)^{L-1} \epsilon_{l} \left[ t_l^{L-1} + (-1)^{L-2} (-\lambda)^{L-2} \epsilon_{r} \right]
              \\
              &= (-\lambda)^{L} + (-1)^{L-1} \epsilon_{l} t_{l}^{L-1} - \epsilon_{l}\epsilon_{r} (-\lambda)^{L-2},
           \end{split}
           \tag{S4}
           \label{S4}
        \end{equation}
     which is the result presented in Eq.~(4) of the main text.

   \section{A physical approach to obtain Eqs.~(6) and (7) }
     In this section, we provide a simplified proof of Eqs.~(6) and (7) in the main text for the case that $H_1$ only contains hoppings between the two ends of the system.

     When the system resides at an IP, the eigenenergy $E_{\rm IP}$ can be viewed as a nonderogatory eigenvalue of the Hamiltonian $H_{\rm IP}$ with multiplicity $k$. Thus, the Hilbert space corresponding to $E_{\rm IP}$ is composed of the eigenvector $|u_0\rangle$ and $k-1$ associated vectors $|u_1\rangle, |u_2\rangle, \cdots, |u_{k-1}\rangle$, which compose a Jordan chain, i.e., 
       \begin{equation}
         \begin{split}
            \left( H_{\rm IP} - E_{\rm IP} \right) &|u_0\rangle = 0,
            \\
            \left( H_{\rm IP} - E_{\rm IP} \right) &|u_1\rangle = |u_0\rangle,
            \\
            &\vdots
            \\
            \left( H_{\rm IP} - E_{\rm IP} \right) &|u_{k-1}\rangle = |u_{k-2}\rangle.
         \end{split}
         \tag{S5}
         \label{S5}
       \end{equation}
     Similarly, the left eigenvector $|v_0\rangle$ corresponding to $E_{\rm IP}$ and the  $k-1$ associated vectors $|v_1\rangle, |v_2\rangle, \cdots, |v_{k-1}\rangle$ compose the left Jordan chain, i.e, 
       \begin{equation}
         \begin{split}
            \langle v_0|& \left( H_{\rm IP} - E_{\rm IP} \right) = 0,  
            \\
            \langle v_1|& \left( H_{\rm IP} - E_{\rm IP} \right) = \langle v_0| ,
            \\
            & \quad \vdots
            \\
            \langle v_{k-1}|& \left( H_{\rm IP} - E_{\rm IP} \right) = \langle v_{k-2}|.
         \end{split}
         \tag{S6}
         \label{S6}
       \end{equation}
     The left and right eigenvectors satisfy the orthogonal normalization condition\cite{seyranian2003multiparameter}
       \begin{equation}
         \langle v_{i}|u_{j}\rangle = \delta_{i+j,k-1} \qquad (i,j = 0,1,\cdots, k-1).
         \tag{S7}
         \label{S7}
       \end{equation}
     It is worth noting that this orthogonal normalization condition differs somewhat from the typical biorthogonal normalization condition used to describe non-Hermitian Hamiltonians whose eigenstates are devoid of defectiveness.

     Now, we add the perturbation $H_{\rm pert} = \epsilon H_1$ and assume that the original eigenenergy $E_{\rm IP}$ is modified as $E_{\rm IP} + \lambda_1$, and the eigenstate $|u_0\rangle$ is modified as
       \begin{equation}
         | u \rangle =  | u_0 \rangle + \sum_{i=1}^{k-1} s_i | u_i \rangle,
         \tag{S8}
         \label{S8}
       \end{equation}
     where $s_i$ denotes coefficients.
     Accordingly,  the Schr\"{o}dinger equation for the perturbed system is 
       \begin{equation}
         \left[ H_{\rm IP} + H_{\rm pert} - E_{\rm IP} - \lambda_1 \right] | u \rangle = 0.
         \tag{S9}
         \label{S9}
       \end{equation}
     Substituting Eqs.\eqref{S5} and \eqref{S8} into Eq.\eqref{S9}, we get  
       \begin{equation}
          \sum_{i=1}^{k-1} s_i | u_{i-1} \rangle + \left( \epsilon H_1 - \lambda_1 \right) \left[| u_0 \rangle + \sum_{i=1}^{k-1} s_i | u_i \rangle \right]  = 0.
          \tag{S10}
          \label{S10}
       \end{equation}
     According to Ref.~\cite{wang2024infernalpoints}, $| u_0 \rangle$ is extremely localized at one side of the system. Considering the normalization condition given in Eq.~\eqref{S7}, it is straightforward to deduce that $| u_0 \rangle$ and $|v_{k-1}\rangle$ are both extremely localized on the same side of the system, while $| u_{k-1} \rangle$ and $|v_{0}\rangle$ are localized on the opposite side. Since $H_1$ only contains hoppings between 
     the two ends of the system, we assume that only two elements of $\langle v_{i}| H_1 |u_{j}\rangle$ are nonzero, which are
       \begin{equation}
         \langle v_{0}| H_1 |u_{0}\rangle = \Lambda, \qquad \langle v_{k-1}| H_1 |u_{k-1}\rangle = \Delta.
         \tag{S11}
         \label{S11}
       \end{equation}
     Using $\langle v_0 |, \langle v_1 | , \cdots, \langle v_{k-1} |$ to multiply Eq.~\eqref{S9} from the left side respectively and utilizing Eq.\eqref{S7}, we obtain
       \begin{equation}
          \begin{split}
            \epsilon \Lambda -\lambda_1 s_{k-1} &= 0 ,
            \\
            s_{k-1} - \lambda_1 s_{k-2} &= 0,
            \\
            &\vdots
            \\
            s_{2} - \lambda_1 s_{1} &=0,
            \\
            s_{1} + \epsilon s_{k-1} \Delta -\lambda_1 &= 0 .
          \end{split}
          \tag{S12}
          \label{S12}
       \end{equation}
      Reducing Eq.~\eqref{S12}, we get
        \begin{gather}
             s_{i} = \lambda_1^{i-1} s_1 , 
             \tag{S13}
             \label{S13}
             \\
             \epsilon \Lambda - \lambda_1^{k-1} s_1 = 0,
             \tag{S14}
             \label{S14}
             \\
             \frac{1}{\lambda_1^{k-1}} \left[ \epsilon \Lambda + \epsilon^2 \Lambda \Delta \lambda_1^{k-2} - \lambda_1^{k} \right] = 0.
             \tag{S15}
            \label{S15}
        \end{gather}
      Since $\epsilon \ll 1$, we neglect the $\epsilon^2$ term in Eq.~\eqref{S15}. Under this approximation, $\lambda_1$ can be solved as
        \begin{equation}
           \lambda_{1,j} \approx \epsilon^{\frac{1}{k}} e_j   \qquad (j=1,2,\cdots,k) ,
           \tag{S16}
           \label{S16}
        \end{equation}
      where $e_j$ for $j=1,2,\cdots,k$ denotes the  $j$-th root of the equation $e - \Lambda^{\frac{1}{k}} = 0$. Substituting Eq.~\eqref{S16} into Eqs.~\eqref{S13} and \eqref{S14}, we get
        \begin{equation}
           s_{i,j} = \left[ \epsilon^{\frac{1}{k}} e_j \right]^i.
           \tag{S17}
           \label{S17}
        \end{equation}
      This means that the single eigenenergy at the IP is split to $k$ eigenenergies, whose expressions are 
        \begin{equation}
           E_{p,j} = E_{\rm IP} + \lambda_{i,j} + o(\epsilon^{\frac{1}{k}}) =  E_{\rm IP} +\epsilon^{\frac{1}{k}} e_j  + o(\epsilon^{\frac{1}{k}}) \qquad  (j=1,2,\cdots,k),
           \tag{S18}
           \label{S18}
        \end{equation}
      and the corresponding eigenstates are
        \begin{equation}
         |u_{p,j}\rangle = |u_0\rangle +  \sum_{i=1}^{k-1 }s_{i,j} |u_i\rangle \approx |u_0\rangle + \epsilon^{\frac{1}{k}} e_j |u_1\rangle \qquad  (j=1,2,\cdots,k).
         \tag{S19}
         \label{S19}
        \end{equation}
      Eq.~\eqref{S18} and Eq.~\eqref{S19} correspond to  Eq.~(6) and Eq.~(7) in the main text, respectively.

      {\wsx

   \section{ Generalized non-Hermitian SSH model as a further example}
      In this section, we employ the generalized non-Hermitian Su-Schrieffer-Heeger (SSH) model 
      {\zb as a further example} to illustrate that the properties of the IP, as presented in the main text, exhibit universality and are independent of the specific Hamiltonian's form.
      \par

      The generalized non-Hermitian SSH model is given by~\cite{wang2024infernalpoints} 
        \begin{equation}
          \begin{split}
            H_{GSSH} = \sum_{n=1}^{L}& (t_1 + \gamma) c_{n,A}^{\dagger} c_{n,B} + (t_1 - \gamma) c_{n,B}^{\dagger} c_{n,A}  \\
            & + (t_2 + l)  c_{n+1,A}^{\dagger} c_{n,B} + (t_2 - l) c_{n-1,B}^{\dagger} c_{n,A}  \\
            & +  (t_3 + \eta)  c_{n-1,A}^{\dagger} c_{n,B} + (t_3 - \eta) c_{n+1,B}^{\dagger} c_{n,A},
          \end{split}
          \tag{S20}
          \label{S20}
        \end{equation}
      where $A$ and $B$ denote the sublattice indices, and  $n \in [1,L]$ labels the position of unit cells. 
      According to Ref.~\cite{wang2024infernalpoints}, this model gives rise to an IP when 
      $t_2 = - l$ and $t_3 = \eta$. At this IP, the system has only two eigenstates,
        \begin{gather}
          | u^{(1)}_0 \rangle = \left( \frac{\sqrt{t_1^2 - \gamma^2}}{t_1 - \gamma},1,0,\cdots,0 \right)^T,
          \tag{S21}
          \label{S21}
          \\
          | u^{(2)}_0 \rangle = \left( -\frac{\sqrt{t_1^2 - \gamma^2}}{t_1 - \gamma},1,0,\cdots,0 \right)^T,
          \tag{S22}
          \label{S22}
        \end{gather}
      with corresponding eigenenergies equal to $E^{(1)}=\sqrt{t_1^2 - \gamma^2}$ and $E^{(2)} = -\sqrt{t_1^2 - \gamma^2}$, respectively. Both states exhibit $L$-fold coalescence. For the convenience of normalization, we set 
        \begin{equation}
           t_3 = \eta = \frac{\sqrt{t_1^2 - \gamma^2}}{2 (t_1 - \gamma)}, \qquad t_2 = -l = \frac{t_1 - \gamma}{2 \sqrt{t_1^2 - \gamma^2}}.
           \tag{S23}
           \label{S23}
        \end{equation}
      For this case, the $(L-1)$-th associated eigenvectors of $| u^{(1)}_0 \rangle$ and $| u^{(2)}_0 \rangle$ are
        \begin{gather}
          |u^{(1)}_{L-1} \rangle = \left( 0,0,\cdots, 0,\frac{\sqrt{t_1^2 - \gamma^2}}{t_1 - \gamma},1 \right)^T,
          \tag{S24}
          \label{S24}
          \\
          |u^{(2)}_{L-1} \rangle = (-1)^{L-1} \left( 0,0,\cdots, 0,-\frac{\sqrt{t_1^2 - \gamma^2}}{t_1 - \gamma},1 \right)^T,
          \tag{S25}
          \label{S25}
        \end{gather}
      respectively. Utilizing the orthogonal normalization condition in Eq.~\eqref{S7}, the corresponding left eigenstates for $| u^{(1)}_0 \rangle$ and $| u^{(2)}_0 \rangle$ can straightforwardly be obtained, which are
        \begin{gather}
          |v^{(1)}_{0} \rangle =\frac{1}{2} \left( 0,0,\cdots, 0,\frac{t_1 - \gamma}{\sqrt{t_1^2 - \gamma^2}},1 \right)^T,
          \tag{S26}
          \label{S26}
          \\
          |v^{(2)}_{0} \rangle = \frac{(-1)^{L-1}}{2} \left( 0,0,\cdots, 0,-\frac{t_1 - \gamma}{\sqrt{t_1^2 - \gamma^2}},1 \right)^T,
          \tag{S27}
          \label{S27}
        \end{gather}
      respectively.
      \par

      We now consider two independent perturbations that connect opposite ends of the system,
        \begin{gather}
           H_{pert}^{(1)} = \frac{\epsilon_1}{2} \left( c_{L,A}^{\dagger} c_{1,A} + c_{L,B}^{\dagger} c_{1,B} +\frac{\sqrt{t_1^2 - \gamma^2}}{t_1 - \gamma}  c_{L,A}^{\dagger} c_{1,B} + \frac{t_1 - \gamma}{\sqrt{t_1^2 - \gamma^2}}  c_{L,B}^{\dagger} c_{1,A} + h.c. \right),
           \tag{S28}
           \label{S28}
           \\
           H_{pert}^{(2)} = \frac{\epsilon_2}{2} \left( c_{L,A}^{\dagger} c_{1,A} + c_{L,B}^{\dagger} c_{1,B} -\frac{\sqrt{t_1^2 - \gamma^2}}{t_1 - \gamma}  c_{L,A}^{\dagger} c_{1,B} - \frac{t_1 - \gamma}{\sqrt{t_1^2 - \gamma^2}}  c_{L,B}^{\dagger} c_{1,A} + h.c. \right) ,
           \tag{S29}
           \label{S29}
        \end{gather} 
      where $\epsilon_1, \epsilon_2 \geqslant 0$ represent the perturbation strengths.
      These perturbations satisfy $\left| \langle v_0^{(1)} | H_{pert}^{(1)} | u_0^{(1)}  \rangle \right| = \epsilon_1$, $\left| \langle v_0^{(2)} | H_{pert}^{(2)} | u_0^{(2)}  \rangle \right| = \epsilon_2$, and $\langle v_0^{(2)} | H_{pert}^{(1)} | u_0^{(1)}  \rangle = \langle v_0^{(1)} | H_{pert}^{(1)} | u_0^{(2)}  \rangle = \langle v_0^{(2)} | H_{pert}^{(2)} | u_0^{(1)}  \rangle = \langle v_0^{(1)} | H_{pert}^{(2)} | u_0^{(2)}  \rangle = 0$. According to our theory, $H_{pert}^{(1)}$ will lead to the splitting of $E^{(1)}$ that scales as $\sqrt[L]{\epsilon_1}$, while $H_{pert}^{(2)}$ similarly affects $E^{(2)}$  with a scaling of $\sqrt[L]{\epsilon_2}$. These predictions are in excellent agreement with the numerical results presented in Fig.~S\ref{figs1a} and Fig.~S\ref{figs1b}.
      \par

      {\zb To demonstrate the robustness of the system's extremely high-order sensitivity to perturbations near---but not exactly at---the IP, we set $t_2 = -l + \delta = \frac{t_1 - \gamma}{2 \sqrt{t_1^2 - \gamma^2}} + \delta$, where $\delta$ quantifies the departure from the exact IP. 
      To further simplify the picture, we set $\epsilon_1 = \epsilon_2 = \epsilon$, so that there is only 
      one parameter to characterize the perturbation strength.
       As shown in Fig.~S\ref{figs1c}, the perturbed spectrum obtained by including $H_{pert}^{(1)} + H_{pert}^{(2)}$ agrees excellently with the 
       predictions from the Eq. (10) of the main text. This not only confirms the robustness of the system's extremely high-order sensitivity to perturbations even under slight deviations from the IP but also establishes that our theoretical framework for the IP's sensitivity is model-independent.}

      \begin{figure}
        \centering
        \subfigure[]{\includegraphics[scale=0.5]{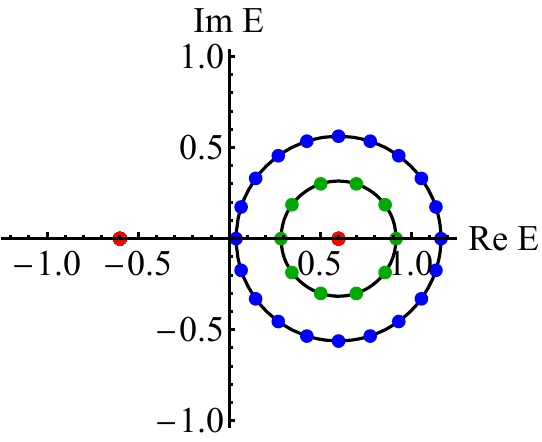} \label{figs1a}}
        \qquad
        \subfigure[]{\includegraphics[scale=0.5]{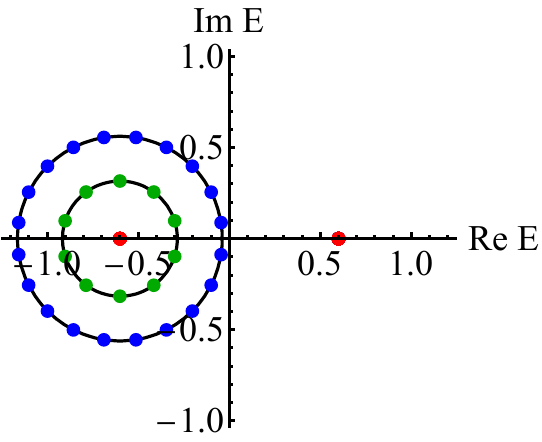} \label{figs1b}}
        \qquad
        \subfigure[]{\includegraphics[scale=0.5]{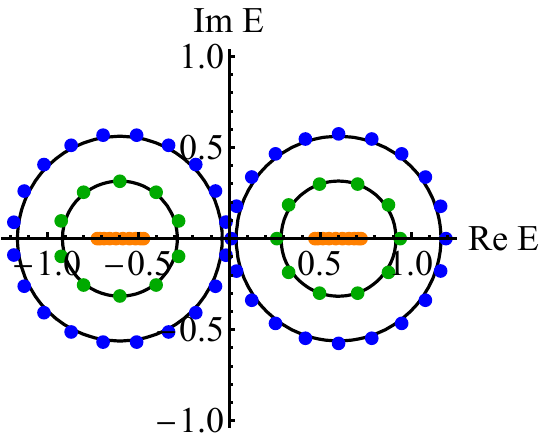} \label{figs1c}}
        \caption{Spectrum for the generalized non-Hermitian SSH model under open boundary conditions. The red points in (a) and (b) are eigenenergies of the system at the IP when $t_2 = - l$ and $t_3 = \eta$. The two red points correspond to $E^{(1)}=\sqrt{t_1^2 - \gamma^2}$ and $E^{(2)} = -\sqrt{t_1^2 - \gamma^2}$. The orange points in (c) show eigenenergies of the system deviated from the IP with $t_2 = -l + \delta$. 
        The green and blue dots refer to spectrum of perturbated system with $L=10$ and $L=20$, respectively.  
        (a) Spectrum for the system at the IP under the perturbation $H_{pert}^{(1)}$ with $\epsilon_1 = 10^{-5}$. The radii for the two circles centered at $E^{(1)}$ are equal to $\sqrt[10]{\epsilon_1}$ (inner) and $\sqrt[20]{\epsilon_1}$ (outer).  (b) Spectrum under the perturbation $H_{\text{pert}}^{(2)}$ with $\epsilon_2 = 10^{-5}$. The radii of the two circles centered at $E^{(2)}$ are equal to $\sqrt[10]{\epsilon_2}$ (inner) and $\sqrt[20]{\epsilon_2}$ (outer). (c) Spectrum under the combined perturbation 
        $H_{\text{pert}}^{(1)} + H_{\text{pert}}^{(2)}$  with $\epsilon_1 = \epsilon_2 =\epsilon = 10^{-5}$. Right circles (centered at $E^{(1)}$) and left circles (centered at $E^{(2)}$) have radii of the value $\sqrt[10]{\epsilon}$ (inner) and $\sqrt[20]{\epsilon}$ (outer). 
        Values of other parameters are $t_1 = 1$, $\gamma = \frac{4}{5}$, $t_3 = \eta = \frac{3}{2}$, $l = -\frac{1}{6}$ and $\delta = \frac{1}{40}$.}
        \label{figs1}
      \end{figure}

      }
\end{document}